\documentclass{ifacconf}

\usepackage{graphicx}
\usepackage{natbib} 
\usepackage{xcolor}
\usepackage{comment}
\usepackage{graphics}
\usepackage{amsmath}
\usepackage{amssymb}
\usepackage{soul}
\usepackage{yfonts}
\usepackage{algorithm}
\usepackage{algpseudocode}
\usepackage[normalem]{ulem}

\makeatletter
\let\old@ssect\@ssect 
\makeatother
\usepackage{hyperref}
\makeatletter
\def\@ssect#1#2#3#4#5#6{%
  \NR@gettitle{#6}
  \old@ssect{#1}{#2}{#3}{#4}{#5}{#6}
}
\makeatother

\def\CD#1{\textcolor{red}{#1}}

\def\hxo{\widehat{x}_0}
\def\d{\mathrm{d}}

\newtheorem{definition}{Definition}
\newtheorem{proposition}{Proposition}
\newtheorem{theorem}{Theorem}

\newtheorem{corollary}{Corollary}
\newtheorem{remark}{Remark}

\begin{document}
\begin{frontmatter}

\title{A scalable, gradient-stable approach to multi-step, nonlinear system identification using first-order methods$^\S$} 

\thanks[footnoteinfo]{corresponding author \textit{cesare.donati@polito.it}.\\
C. Donati acknowledges support from PRIN project TECHIE “A control and network-based approach for fostering the adoption of new technologies in the ecological transition”, Cod. 2022KPHA24 CUP: D53D23001320006.\\$^\S$This work has been submitted to IFAC for possible publication.}

\author[DET,CNR]{Cesare Donati\thanksref{footnoteinfo}}
\author[CNR]{Martina Mammarella} 
\author[CNR]{Fabrizio Dabbene}
\author[DET]{Carlo Novara}
\author[PSU]{Constantino Lagoa}

\address[DET]{DET, Politecnico di Torino, Torino, Italy {(e-mail: cesare.donati@polito.it, carlo.novara@polito.it)}}%
\address[CNR]{Cnr-Istituto di Elettronica e di Ingegneria dell'Informazione e delle Telecomunicazioni, c/o Politecnico di Torino, Torino, Italy {(e-mail: martina.mammarella@cnr.it, fabrizio.dabbene@cnr.it)}}%
\address[PSU]{EECS, The Pennsylvania State University, University Park, PA, USA {(e-mail: cml18@psu.edu)}}

\begin{abstract}
This paper presents three main contributions to the field of multi-step system identification. First, drawing inspiration from Neural Network (NN) training, it introduces a tool for solving identification problems by leveraging first-order optimization and Automatic Differentiation (AD). The proposed method exploits gradients with respect to the parameters to be identified and leverages Linear Parameter-Varying (LPV) sensitivity equations to model gradient evolution. Second, it demonstrates that the computational complexity of the proposed method is linear in both the multi-step horizon length and the parameter size, ensuring scalability for large identification problems. Third, it formally addresses the ``exploding gradient" issue: via a stability analysis of the LPV equations, it derives conditions for a reliable and efficient optimization and identification process for dynamical systems. Simulation results indicate that the proposed method is both effective and efficient, making it a promising tool for future research and applications in nonlinear system identification and non-convex optimization.
\end{abstract}


\end{frontmatter}

\section{Introduction}
Over the past few decades, the field of nonlinear system identification has received significant attention. Modern engineering applications increasingly require sophisticated models, capable of delivering accurate estimates over extended time periods, highlighting the need for approaches that ensure the minimization of multi-step prediction errors {arising from the propagation of predictions over multiple time steps}. {Indeed, single-step techniques{, while effective for immediate prediction,} may lack accuracy in multi-step prediction and fail to capture the system's relevant dynamics} \citep{terzi2018learning,farina2011simulation}.
Clearly, the identification of such types of models may easily
destroy nice convexity properties of the associated cost functions, thus leading to hard optimization problems \citep{terzi2018learning}. {This is primarily due to the nonlinear interactions between parameters, which become increasingly complex as predictions propagate over longer horizons}.
In such a situation, first-order methods offer a practical and effective solution. 
Such methods, which under suitable conditions are guaranteed to converge to a solution (in general sub-optimal), leverage gradient information to iteratively update decision variables \citep{nesterov2018lectures} and have recently gained popularity for their ability to tackle large-scale and complex problems~\citep{ahookhosh2019accelerated, teboulle2018simplified}. Indeed, their effectiveness in solving non-convex problems
remains one of the unresolved mysteries, contributing to the success of deep learning across numerous applications (see \cite{min2023convergence} and references therein), and
makes them an appealing choice for tackling the optimization challenges presented by complex identification tasks. 

In this work, we present a {computable tool} to leverage first-order methods, which have been at the core of the success of NN-based approaches, for solving optimization problems central to multi-step identification, i.e., aimed at minimizing prediction errors over multiple steps ahead in time. 
In particular, we exploit the explicit knowledge of the physical dependencies characterizing the system to identify, {such as known physical laws, constraints, and relationships between system variables,}
and propose an efficient method to compute the gradient and estimate system parameters. 

Certainly, various standard methods are available in the literature for calculating the gradient, with varying degrees of approximation, and, at least in principle, all of these can be applied, ranging from techniques based on numerical differentiation to generic automatic differentiation and classical backpropagation \citep{baydin2018autodiffsurvey,margossian2019reviewAD}. 
In the context of system identification, some works have recently explored the application of classical backpropagation to parameter estimation (see, e.g., \cite{kaheman2022automatic,donati2023oneshot,FerrariTrecate2023simba}).
However, 
a more in-depth analysis of the application of automatic differentiation to multi-step identification problems and, in particular, to the framework we propose in \cite{donati2024automatica},
sheds light on several distinguishing characteristics that make it of particular interest, as shown by the three main contributions of this work.
First, the proposed method leverages a ``smart'' and efficient way based on cost sensitivity equations to evaluate the \textit{exact} gradient when dealing with the identification of dynamical systems.
{Specifically, a direct application of AD and the chain rule to the cost function allows the definition of an LPV dynamical system describing the evolution {along the prediction horizon} of the gradient with respect to model parameters and the initial condition, directly linked to the dynamics of the system we aim at identifying (see, e.g., \cite[Appendix A]{ribeiro2020smoothness}, \cite{pillonetto2025deep})}. 
Second, it is demonstrated to be well suited for multi-step identification, exhibiting a computational complexity linear in both the multi-step horizon length and the parameter size. {These are crucial factors in this context, as an increase in either can cause the computational complexity to grow rapidly if not managed by proper algorithms \citep{schoukens2019nonlinear}.}
Third, a stability study on the LPV gradient dynamics allows us to derive conditions for avoiding the so-called ``exploding gradient" phenomenon when identifying dynamical systems. 
Indeed, in the context of first-order methods, it is crucial to ensure the stability of the gradients used for optimization. 
For instance, model predictions may exhibit finite escape 
time phenomena for certain identified parameters or initial condition values, leading to divergent behavior and consequent exploding gradients. This behavior might hinder the identification process,
highlighting
the necessity of avoiding unstable predictions
and remaining in a ``safe" neighborhood around the measured trajectory, while ensuring a consistent and reliable optimization process. 
Here, it is important to emphasize that the gradient stability analysis based on the LPV dynamical system cannot rely solely on the eigenvalues of the pointwise-in-time state-transition matrix. 
Indeed, differently from linear time-invariant systems, eigenvalues of time-varying systems provide only a limited, static perspective and cannot fully characterize the stability of an LPV system\footnote{For a detailed discussion on the limitations of eigenvalue-based analysis in time-varying systems, see \cite[Chapter 24]{rugh1996linear}.}.
For this reason, our approach considers the Bounded-Input, Bounded-Output (BIBO) stability of the entire gradient dynamics over the multi-step horizon. This broader perspective ensures conditions such that the evolving gradient remains bounded
.

The remainder of this paper is structured as follows. In Section~\ref{sec:recap} the framework presented in \cite{donati2024automatica} is summarized, analyzing multi-step identification problems. In Section \ref{sec:RGC} the use of AD in system identification is discussed, detailing the proposed gradient dynamics, while in Section \ref{sec:compcompl} the computational complexity of the proposed method is studied. In Section \ref{sec:explgrad} the stability of the gradient dynamics is analyzed, while numerical results are discussed in Section \ref{sec:numresults}. Main conclusions are drawn in {Section~\ref{sec:concl}}.

\textit{Notation.}
Given integers ${a,b\in \mathbb N}$, ${a\leq b}$, we denote by $[a,b]$ the set of integers $\{a,a+1,\ldots,b\}$.
The symbol $\nabla$  denotes the gradient operator, where  $\nabla_x L$  represents the vector of partial derivatives of $L$ with respect to $x$. The Jacobian matrix of~${v_k \in \mathbb R^{n_v}}$ with respect to $w_k \in \mathbb R^{n_w}$ is denoted as~$\mathcal{J}^{v\!/\!w}_k~\in~\mathbb R^{n_v, n_w}$,
i.e., $\frac{\partial v_k}{\partial w_k}$. Similarly, $\mathcal{J}^{v\!/\!v}_k \in \mathbb R^{n_v, n_v}$ is the Jacobian matrix of $v_k$ with respect to $v_{k-1}$, i.e.,~$\frac{\partial v_k}{\partial v_{k-1}}$.

\section{ Preliminaries}\label{sec:recap} 
\subsection{Problem setup}
In this section, we {briefly} outline the multi-step system identification problem, recalling the framework presented in \cite{donati2024automatica}.
%
%
We consider a nonlinear, time-invariant system described by the following mathematical model
\begin{equation}
    \begin{aligned}
    \mathcal{S}:\quad&x_{k+1} = f\left({x}_k, {u}_k; \theta\right) + \Delta(x_k,u_k),\\
    &z_k = h\left(x_k\right),    \end{aligned}
\label{eqn:system}
\end{equation}
where $x \in \mathbb{R}^{n_x}$ is the state vector, $\theta \in \mathbb{R}^{n_\theta}$ is the parameter vector,
$u \in \mathbb{R}^{n_u}$ is the input, and
$z \in \mathbb{R}^{n_z}$ is the output.
Functions $f$ and $h$ are known and assumed to be nonlinear, time-invariant, and continuously differentiable. The term $\Delta$, representing unmodeled dynamics, is unknown.

Given a multi-step input sequence $\widetilde{\mathbf{u}}_{0:T} = \{\widetilde u_0,\dots, \widetilde u_{T}\}$, $\widetilde u_k = u_k + \eta^u_k$, and the corresponding observations sequence $\widetilde{\mathbf{z}}_{0:T} = \{\widetilde z_0,\dots, \widetilde z_{T}\}$, $\widetilde z_k = z_k + \eta^z_k$,
with $\eta^u_k$ and $\eta^z_k$ {being} the input and output measurement noise, we seek to estimate the unknown system parameters $\theta$, and initial conditions $x_0$, while compensating for the unknown term $\Delta$.
To this aim, we define an estimation model $\mathcal{M}$ to approximate the system $\mathcal{S}$, i.e.,
\begin{equation}
\begin{aligned}
{\mathcal{M}}\,:\quad\widehat{x}_{k+1} &= {f}(\widehat{x}_{k},  {u}_{k}; \widehat \theta) + {\delta}(\widehat{x}_{k},u_k; \omega),\\
    \widehat {{z}}_{k} &= {h}(\widehat{x}_{k}),
\label{eqn:bb_extension_v2}
    \end{aligned}
\end{equation}
where $\delta$ is a generic approximator (e.g., a linear combination of basis functions from a given dictionary) with parameters $\omega \in \mathbb R^{n_\omega}$ to be learned.

In the proposed framework, considering the multi-step prediction error at time $k$, i.e., $e_k = \widetilde z_k - \widehat z_k$, the multi-step cost function takes the form
\begin{equation}
\mathcal{C}_T(\theta,x_0,\omega; \mathbf{e}_{0:T},\widehat{\mathbf{x}}_{1:T})\doteq
    \sum_{k=0}^{T} \mathcal{L}_k\left(\theta,x_0,\omega;e_k,\widehat x_k\right),
    \label{eqn:final_cost}
\end{equation}
where $\mathcal{L}_k$ is a general, twice continuously differentiable loss function, including possible {physics-based} penalties and regularization terms \citep{donati2024automatica,mammarella2024blended}. 
The final goal is to estimate the optimal values of $\theta$, $x_0$, and $\omega$ by solving the following optimization problem
\begin{equation}
    \left( {\theta}^\star, x_0^\star, \omega^\star\right) \doteq \arg \min_{\theta, x_0,\omega}\,\, \mathcal{C}_T.
    \label{eqn:optprobl}
\end{equation}
The interested reader is referred to \cite{donati2024automatica} for additional details on the framework.
\subsection{Multi-step identification}
\begin{figure}[!tb]
    \centering
    \includegraphics[trim = {4cm 4.5cm 4cm 1.21cm},clip,width = 0.85\columnwidth]{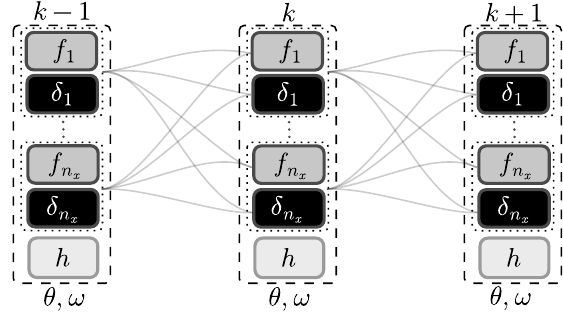}
    \caption{Multi-step model propagation.}
    \label{fig:NNlike}
\end{figure}

Multi-step identification involves minimizing a multi-step cost function over a horizon $T$, propagating the predictions~${\widehat x_k}$ over the desired horizon, and recursively applying the dynamical model $\mathcal{M}${, evaluating the multi-step
prediction error $e_k$}. 
A somewhat similar approach is found in RNNs \citep{scattoliniRNN}, which aim at accurately estimating an output sequence by minimizing the sum-of-squares error measure over a finite horizon, involving the recursion of the hidden states repeatedly through the same layer. This implies that the network weights, as well as the architectures, are the same over the entire horizon. 
Thus, recalling the typical structure of neural networks, it is possible to represent the multi-step propagation of the model \eqref{eqn:bb_extension_v2} as in Fig.~\ref{fig:NNlike}, drawing an analogy between RNN layers and the dynamical model propagation. Notice that this representation is very general, and can be used for \textit{any} dynamical system of the form \eqref{eqn:bb_extension_v2}.
As in RNNs, during the identification process, we need to consider the weights, i.e., parameters to identify, and the structure, i.e., functional form of the model, to be the same at {each time step} along the prediction horizon $T$, in order to preserve physical consistency and temporal dependencies in the learning process. 
This similarity led us to investigate techniques commonly used for optimizing recurrent neural networks, such as backpropagation through time \citep{donati2023oneshot, TBTT}, and, more in general, automatic differentiation, in the context of multi-step system identification.

In the next section, we focus on analyzing how the gradient may be efficiently evaluated in multi-step identification problems. 
Thus, a (possibly local) solution for the optimization problem~\eqref{eqn:optprobl} can be obtained through first-order methods. 

\section{Automatic differentiation in multi-step system identification} \label{sec:RGC}
At the core of first-order techniques is the computation of the gradient of the cost function $\mathcal{C}_T$ evaluated at the current solution. 
%
In this context, automatic differentiation is a general computational technique used to efficiently compute the gradients of functions, often applied in optimization and machine learning. It relies on the concept of graph computation, where a function is decomposed into a series of elementary operations represented as nodes, with edges indicating data flow between them. Thus, AD systematically applies the chain rule of calculus to propagate derivatives through this graph. 

In the context of multi-step identification of dynamical systems, the computational graph becomes very specialized due to the recursive nature of the problem. This recursive structure allows for efficient gradient evaluation at each time step, effectively turning the gradient evolution into a dynamical system itself, directly connected to the estimation model~\eqref{eqn:bb_extension_v2}, 
and making the process both general and well-suited for the identification of any dynamical system.

In this section,
we exploit the explicit knowledge of the structure of the dynamical system~\eqref{eqn:system} and the defined estimation model \eqref{eqn:bb_extension_v2}, and present an efficient adaptation of the forward automatic differentiation technique to the context of system identification.
\begin{remark}[on the gradient LPV equations]
Within this context, we remark that the following formulation is not new and has appeared in the literature under various forms and names, such as adjoint sensitivity analysis and state sensitivity equations (see, e.g., \citep{ribeiro2020smoothness, pillonetto2025deep}). In this work, we present a tailored version designed for the identification of nonlinear dynamical systems. This formulation will be exploited in Section \ref{sec:explgrad} to perform the stability analysis of the gradient, providing insights to ensure its boundedness during the optimization process.
\end{remark}

\subsection{Gradient LPV equations}\label{sec:Gdyn}
%
The following results are presented considering a cost function of the form \eqref{eqn:final_cost}.
Specifically, in the following proposition, whose proof is reported in Appendix \ref{app:prop1}, we show how to represent the evolution of the gradient with respect to $\theta$ as a dynamical system.
\begin{proposition}[gradient dynamics -- $\theta$]
\label{prop1}
{Define the \textit{memory matrix}}
\begin{equation}
\label{eqn:Lambdadef}
\Lambda_{k} \doteq \frac{\d {x_k}}{\d {\theta}} \in \mathbb R^{n_x,n_\theta},
\end{equation}
as the matrix containing the total derivatives of the states with respect to $\theta$.
Define  vectors $\rho_k \in \mathbb R^{n_x}$, $\varrho_k \in \mathbb R^{n_\theta}$ as 
\begin{equation}
    \label{eqn:rho-var-def}
    \rho_k \doteq \left({\nabla_{e}^\top  \mathcal{L}_k}{\mathcal{J}^{e\!/\!z}_k}{\mathcal{J}^{z\!/\!x}_k}\right)^\top,\quad
    \varrho_k \doteq {\nabla_{\theta} \mathcal{L}_k}.
\end{equation}
{Then, }the gradient evolution with respect to the parameter $\theta$ over the multi-step horizon $T$ is described by the following time-varying dynamical system
\begin{subequations}
    \begin{align}
     \label{eqn:GUlaw_mat}
     &\Lambda_{k} = \mathcal{J}^{x\!/\!x}_k\Lambda_{k-1}+
     \mathcal{J}^{x\!/\!\theta}_k, \\
     \label{eqn:GUlaw_grad}
     &\nabla_{{\theta}} \mathcal C_{k} = \nabla_{{\theta}} \mathcal C_{k-1} + \Lambda^\top_k\rho_k + \varrho_k,    
    \end{align}
    \label{eqn:GULAW}%
\end{subequations}
for $k=1,\dots,T$, with
    $\Lambda_{0} \doteq \frac{\d x_0}{\d \theta} = \mathbf{0}_{n_x,n_\theta}$, $\nabla_{{\theta}} \mathcal C_{0} = \mathbf{0}_{n_\theta}$. 
\end{proposition}

In \eqref{eqn:GUlaw_mat} the term $\mathcal{J}_k^{x\!/\!\theta}$ reflects the direct effect of the model parameter estimation $\widehat \theta$ on the current state prediction $\widehat x_k$ and, consequently, on $\mathcal{L}_k$. Conversely, the term $\Lambda_{k-1}$ encapsulates how the effect of $\widehat \theta$ on past predictions, i.e.,~$\widehat x_\tau$ with $\tau\in[1,k-1]$, has affected the current state estimation $\widehat x_k$.
Thus, \eqref{eqn:GUlaw_grad} defines a formula in which the gradient is updated at each time step $k$ with an \textit{innovation term}, i.e., $\Lambda^\top_k\rho_k + \varrho_k$, exploiting the information encapsulated in $\Lambda_{k}$. 
Here, notice that $\mathcal{J}^{e\!/\!z}_k$, $\mathcal{J}^{z\!/\!x}_k$, $\mathcal{J}^{x\!/\!x}_k$, $\mathcal{J}^{x\!/\!\theta}_k$ are time-varying matrices with fixed structure and shape, only depending on the values of~$\widehat x_k, \widetilde u_k, \widehat z_k, \widetilde z_k, e_k$, $\widehat \theta$, $\hxo$, and $\omega$ at time-instant $k$.
For instance, we have that
$$
    \mathcal{J}^{x\!/\!x}_k = \frac{\partial \widehat x_k}{\partial \widehat x_{k-1}} = \frac{\partial \left(f(\widehat x_{k-1}, \widetilde u_{k-1}; \widehat \theta)\!+\!{\delta}\left(\widehat{x}_{k-1},\widetilde u_{k-1}; \omega\right)\right)}{\partial \widehat x_{k-1}}.
$$

While an analogous result can be obtained considering the extended parameter vector~${\vartheta = [\theta^\top,\omega^\top]^\top \in \mathbb R^{n_\theta+n_\omega}}$, a slightly different dynamic is obtained in the case of the gradient with respect to the initial condition.
This is shown in the following proposition, whose proof follows the same reasoning of the proof of Proposition \ref{prop1}, and it is
reported in Appendix \ref{app:prop2}.
\begin{proposition}[gradient dynamics -- $x_0$]
\label{prop2} Let
\begin{equation}
    \Lambda_{0,k} \doteq \frac{\d x_k}{\d {x}_{0}} \in \mathbb R^{n_x,n_x}
    \label{eqn:phi0def}
\end{equation}
be the matrix containing the total derivatives of the states with respect to $x_0$. Consider \eqref{eqn:rho-var-def}.
The gradient evolution with respect to the initial condition along the multi-step horizon $T$ is obtained by means of the following time-varying dynamical system
\begin{subequations}
    \begin{align}
        \label{eqn:GUlaw_initcond_mat}
         &\Lambda_{0,k} = \mathcal{J}^{x\!/\!x}_k\Lambda_{0,k-1},\\
       \label{eqn:GUlaw_initcond}
         &\nabla_{x_0} \mathcal C_{k} = \nabla_{x_0} \mathcal C_{k-1} + {\Lambda^\top_{0,k}}\rho_k,
    \end{align} 
    \label{GULAW_x0}%
    \end{subequations}
with $\Lambda_{0,0} \doteq \frac{\d x_0}{\d x_0} = \mathbb{I}_{n_x}$, 
$\nabla_{x_0} \mathcal C_{0} = {\Lambda^\top_{0,0}}\rho_0$. 
\end{proposition}

{In this case, there is no ``direct" effect of $\widehat x_0$ on~$\widehat{x}_k$, since, differently from $\widehat \theta$, the estimated initial condition does not affect future predictions entering ``directly" into the model at each time step}, but only through its propagation over time. This effect is captured by the ``memory" term $\Lambda_{0}$.

%

\subsection{Proposed approach}\label{sec:propapproach}
Based on the results of Propositions \ref{prop1} and \ref{prop2}, we outline in Algorithm~\ref{alg:algGD} the proposed optimization approach for multi-step identification. We propagate the model~\eqref{eqn:bb_extension_v2} with initial conditions $\widehat x_0$, model parameters $\widehat\theta$, and $\omega$, while updating the gradient according to Proposition \ref{prop1} and \ref{prop2} along the multi-step horizon $T$. Then, accordingly, we update the weights. 
This process repeats until at least one of the following conditions is satisfied: i) the optimization converges to a (possibly local) minimum of the loss function, or below a given threshold~$\varepsilon_1$; ii) the magnitude of the gradient is lower than a given minimum threshold~$\varepsilon_2$.
\begin{algorithm}
\caption{First-order identification algorithm}\label{alg:algGD}
\begin{algorithmic}[1]
\State Given $\{\mathbf{\widetilde u}_{0:T},\mathbf{\widetilde z}_{0:T}\}$, choose $\delta(\cdot)$ \eqref{eqn:bb_extension_v2}, $\varepsilon_1$, and $\varepsilon_2$.
\State {Initialize} $\widehat{x}_0$, $\widehat\vartheta = [\widehat\theta^{\top},\omega^{\top}]^\top$.
\While{$\mathcal{C}_T \geq \varepsilon_1$ \textbf{and} $\|\nabla\mathcal{C}_T\|_2 \geq \varepsilon_2$ }
\State {Initialize} $k = 0$, $\Lambda_{0} = \mathbf{0}_{n_x,n_\theta}$, $\nabla_{{\vartheta}} \mathcal C_{0} = \mathbf{0}_{n_\vartheta}$, $\Lambda_{0,0} = \mathbb{I}_{n_x}$, and $\nabla_{x_0} \mathcal C_{0} = {\Lambda^\top_{0,0}}\rho_0$.
\While{ $k \leq T$ }
\State Predict $\widehat x_{k+1}$, $\widehat z_k$ using \eqref{eqn:bb_extension_v2}, with $\widehat{x}_0$, $\widehat{\vartheta}$.
\State Compute ${e}_{k} = \widehat z_k-\widetilde z_k$ and $\mathcal{L}_k$.
\State Compute $\rho_{k}$ and $\varrho_k$  using \eqref{eqn:rho-var-def}.
\State Compute $\Lambda_{k}$ \eqref{eqn:GUlaw_mat} and $\Lambda_{0,k}$ \eqref{eqn:GUlaw_initcond_mat}.
\State Compute  $\nabla_{{\vartheta}} \mathcal C_{k}$ and $\nabla_{x_0} \mathcal C_{k}$ using  \eqref{eqn:GUlaw_grad} and \eqref{eqn:GUlaw_initcond}.
\State $k \gets k+1$.
\EndWhile
\State Compute $\mathcal{C}_T$ using \eqref{eqn:final_cost}.
\State Define $\nabla\mathcal{C}_T = [\nabla^\top_\vartheta\mathcal{C}_T, \nabla^\top_{x_0}\mathcal{C}_T]^\top$.
\State Update $\widehat \vartheta, \widehat x_0$ using any first-order method.
\EndWhile
\State Return $\vartheta^\star=\widehat\vartheta$ and ${x}_0^\star=\widehat{x}_0$
\end{algorithmic}
\end{algorithm}

\section{Computational complexity}\label{sec:compcompl}
In the context of multi-step system identification, the computational complexity can easily explode as the multi-step horizon and parameter size increase \citep{schoukens2019nonlinear}. Therefore, it is crucial to have a reliable algorithm whose complexity does not grow exponentially with these values.
Thus, with the following theorem, whose proof is reported in Appendix \ref{app:compl_improv}, we formally state the computational complexity of the proposed gradient computation algorithm for a given multi-step horizon and parameter size. As a measure of computational complexity, we consider the maximum number of operations needed to execute a given algorithm, expressed in Big O notation.
\begin{theorem}[complexity analysis]\label{rmk:compl_improv}
    Let $T$ be the length of the multi-step horizon considered in the system identification process, $n_x$ the number of states, and $n_\vartheta = n_\theta+n_\omega$ the total number of parameters in \eqref{eqn:bb_extension_v2}. The computational complexity required to compute the gradient by iterating the dynamical system~\eqref{eqn:GULAW} scales linearly with the length of the multi-step horizon $T$ and the parameter size, exhibiting a computational complexity of $\mathcal{O} (Tn_x^2n_\vartheta)$. 
\end{theorem}

Note that, while $n_x$ is typically fixed and depends on the system \eqref{eqn:system} under analysis, $T$ and $n_\vartheta$ may vary, depending on the specific requirements of the problem. For instance, $n_\vartheta$ can easily increase, as it depends on the size of $\omega$ for the selected black-box model~$\delta$.

It is important to remark that, while \eqref{eqn:GULAW} and \eqref{GULAW_x0} result from the application of forward AD in the context of multi-step system identification, a similar result can be obtained via the application of backward AD, as discussed {next}.
\begin{remark}[backward AD in system identification]\label{rmk:bkwd_ad}~ 
Forward AD provides a more intuitive and flexible representation of the gradient as it evolves forward in time with the model predictions, which enables the stability analysis in Section~\ref{sec:explgrad}. In contrast, backward AD requires processing the entire horizon before the gradient can be computed via reverse-time adjoint calculations \citep{baydin2018autodiffsurvey}. 
Although this result is not included in this paper, we note for completeness that the computational complexity of the gradient computation via backward AD is $\mathcal{O}(T(n_x^2 + n_\vartheta n_x))$, and it can be derived using a reasoning similar to the one reported in Appendix \ref{app:compl_improv}.
Therefore, we remark that backward AD offers a valid alternative and a complementary perspective on the gradient dynamics, and is generally more efficient for high-dimensional problems with many parameters to identify, such as in the case of large black-box models $\delta$.
\end{remark}

Lastly, we also highlight that the proposed approach offers a more efficient way to compute the gradients compared to the analytic formula proposed in \cite{donati2023oneshot}, as stated in the following remark and numerically illustrated in Section \ref{sec:computanalitically}.  
\begin{remark}[complexity improvement]\label{rmk:compl_comp}
    Following a reasoning similar
 to the one reported in Appendix \ref{app:compl_improv}, it can be proved that the computational complexity of the algorithm proposed in \cite{donati2023oneshot} {scales as $\mathcal{O}(T^3n_x^2 + T^2n_\theta)$, exhibiting a cubic dependency with respect to the multi-step horizon $T$.} 
\end{remark}

\section{Non-exploding gradient}
\label{sec:explgrad}
In the context of first-order methods, it is important to ensure the stability of the gradients used for the optimization. 
In this section, we investigate this issue by exploiting the dynamical formulation \eqref{eqn:GULAW} to analyze the stability of the gradient, useful to ensure that gradients do not grow unboundedly and ensure a reliable identification. {In this context, we offer a comprehensive approach grounded in systems theory to ensure an efficient stability characterization of the gradient dynamics. }
While the analysis is carried out for $\theta$, we remark that the same reasoning with analogous results also applies to the gradient for the initial condition.
First, the concept of non-exploding (or stable) gradient is formally defined as follows. 
\begin{definition}[non-exploding gradient]\label{def:nexpgrad}~
    The multi-step gradient $\nabla_{{\theta}} \mathcal C_{T}$ is said to be non-exploding (or stable) if and only if there exists a finite constant $\mu_1$ such that it satisfies
    \begin{equation}
    \sup_{k \in [0, T]} \|\nabla_{{\theta}} \mathcal C_{k}\|_2 \leq \mu_1.
    \label{eqn:nexpdef}
    \end{equation}
\end{definition}
Definition \ref{def:nexpgrad} follows from basic stability notions. For the gradient to be classified as non-exploding, it must have a finite ``gain” throughout its evolution along the multi-step horizon~$T$, described by \eqref{eqn:GULAW}.
Thus, it is possible to establish a more specific condition for gradient stability by formalizing the link between multi-step gradient dynamics defined by the LPV system \eqref{eqn:GULAW} and BIBO stability
(see e.g., \cite{rugh1996linear}), as stated with the following theorem, whose proof is reported in Appendix~\ref{app:exp.grad}.
\begin{theorem}[gradient stability] \label{thm:exp.grad}
    The multi-step gradient $\nabla_{{\theta}} \mathcal C_{T}$ is non-exploding according to Definition \ref{def:nexpgrad} if and only if the associated LPV system \eqref{eqn:GULAW} is BIBO stable, i.e., there exists a finite constant $\mu_2$ such that
    \begin{equation}
    \sum_{i=j}^{k-1}\|\Lambda^\top_i\rho_i+ \varrho_i\|_2\leq \mu_2,
    \label{eqn:theo2cond}
    \end{equation}
    for all $k \in [1, T]$,  $j$ with $k\geq j+1$.  
\end{theorem}

Consequently, it follows from Theorem \ref{thm:exp.grad} and \eqref{eqn:GUlaw_mat} that certain conditions on the system's predicted trajectory can provide sufficient guarantees for the gradient to be bounded over the multi-step horizon $T$, as stated in the following corollary, proved and discussed in Appendix \ref{app:corollary.exp.grad}.
\begin{corollary}[trajectory to gradient stability]\label{cor:corollary.exp.grad}
    Let $k_c<\infty$ be the critical time step at which a given trajectory diverges to infinity, such that
    $ \lim_{k \to k_c} \lVert x_k \rVert_p = \infty$, for a suitable $\ell_p$ norm.
    For a specific current value of $\widehat \theta \in \mathbb R^{n_\theta}$, a sufficient condition for the gradient to be stable is that the predicted trajectory defined by the sequence of states $\{\widehat x_0,\dots\widehat x_T\}$ satisfying the model dynamics \eqref{eqn:bb_extension_v2} does not have a finite escape time $k_c \leq T$, 
   i.e., it is not \textit{finite-time unstable} with $k_c \leq T$.  
\end{corollary}

Thus, a simple yet effective consequence of Corollary \ref{cor:corollary.exp.grad} is that if there exist some physical insights regarding the stability of system~\eqref{eqn:system}, 
then it is possible to properly add penalty terms in \eqref{eqn:final_cost} in order to maintain a stable gradient during the identification process. 

Clearly, knowing the parameter values that lead to instability would be beneficial, as this enables directly imposing constraints on the parameter values. However, it should be remarked that, in most cases, such values are not known to the user. In general, 
only the values of the states at which the system does not show instability are known, leading to state-dependent penalty terms, as detailed in the following remark.
\begin{remark}[trajectory stability via state barriers]\label{rmk:state_barriers}~
Assume that the underlying system is known to be stable within the interval
$
    x \in \mathcal{X} \doteq \left\{x^{lb}_i \leq \widehat x_i \leq x^{ub}_i,\,\, i = [1,n_x]\right\}.
$ 
Then, the optimization can be steered in order to remain in this safe area by expressing in \eqref{eqn:final_cost} a penalty term defined by the composition of two barrier functions, i.e.,
\begin{equation}
    p(\widehat{x}_k, \theta) \doteq \lambda \|e^{\alpha(\hat{x}_k - x^{ub})}\|_2^2 + \lambda\|e^{\alpha(x^{lb} - \hat{x}_k)}\|_2^2,
    \label{eqn:penalty_term}
\end{equation} with $\lambda, \alpha\in\mathbb R$ tunable parameters. 
\end{remark}
As a consequence, the predicted state variables are encouraged to stay within the specific intervals where the trajectories are known to be stable, as shown in the example proposed in Section \ref{sec:expl_grad_ex}.
In this context, it is important to highlight that the problem of non-exploding gradient can still be handled by relying on other standard techniques, such as gradient clipping or truncated gradient (see, e.g., \cite{TBTT} and references therein). However, in this case, the obtained gradient may introduce higher errors, being itself an approximated version of the exact one. 

\section{Numerical example}\label{sec:numresults}
In this section, two numerical examples are provided. First, a comparison with \cite{donati2023oneshot} in terms of computational complexity is proposed. Then, the results of Theorem~\ref{thm:exp.grad} are shown through simulation for the identification of the parameters of a population dynamics model.

\subsection{Complexity comparison }\label{sec:computanalitically}
The improvement in computational complexity described in Remark \ref{rmk:compl_comp} is illustrated by briefly revisiting the same numerical example proposed in \cite{donati2023oneshot}, in which the attitude dynamics of a satellite {was} modeled using the Euler equations. {The simulations are run in MATLAB 2021b over an Apple M1 with an 8-core CPU, 8GB of RAM, and a 256 GB SSD unit.} Both methods yield identical gradients in terms of accuracy and values, however, as shown in Table \ref{tab:comp_rec}, by dynamically computing the gradient using \eqref{eqn:GULAW}, the computational efficiency is significantly improved, making the optimization both faster, and consequently significantly more scalable. Additionally, it can be observed that the computational times follow the linear and {cubic} trends with respect to~$T$ stated in Theorem \ref{rmk:compl_improv} and Remark \ref{rmk:compl_comp}, respectively.
\begin{table}[!tb]
\centering
\caption{Average gradient calculation time (seconds) for different values of $T$ ($n_x=3$, $n_\theta = 3$).}
\begin{tabular}{c c c}
    \hline
    T & \eqref{eqn:GULAW} & \cite{donati2023oneshot} \\
    \hline
    50  & $0.0015$ & $0.0411$ \\
    100 & $0.0020$ & $0.2877$ \\
    150 & $0.0035$ & $0.9300$ \\
    200 & $0.0043$ & $2.3048$ \\
    250 & $0.0049$ & $4.2128$ \\
    \hline
\end{tabular}
\label{tab:comp_rec}
\end{table}

\subsection{Exploding gradient: a population dynamics example}\label{sec:expl_grad_ex}
Let us consider a population dynamics model described by the discrete-time logistic map \citep{logistic_map_1, may1976simple}, i.e.,
$$
    x_{k+1} = \theta x_k (1-x_k),
$$
where $x_k\in \mathbb R$ represents the ratio of the existing population to the maximum possible population at time step $k$, restricted to the interval $[0,1]$, and $\theta\in \mathbb R$ is the parameter representing the growth rate.
The behavior of the logistic map depends crucially on $\theta$. In particular, while for ${0\leq \theta \leq 4}$ we have that $x_k$ converges, oscillates, or exhibits chaotic behavior in $[0,1]$, for $\theta>4$ it leaves the ``safe" interval $[0,1]$ and diverges in finite-time, for almost all initial conditions. 

{In this example, we aim at identifying the growth rate $\theta$ that characterizes a given, stable, nominal population evolution over a horizon $T=10^4$, i.e.,  $\{\widetilde x_0, \dots, \widetilde x_T\}$. }
Since the state variable represents the ratio between the current and maximum population, it is reasonable to assume that the values of the states where the system exhibits stable behavior are available. Thus, according to Remark \ref{rmk:state_barriers}, it is possible to rely on a penalty term of the form \eqref{eqn:penalty_term} to bind the 
predicted trajectories in the desired interval, i.e.,~$[0,1]$, thus ensuring stable gradients and allowing the identification of $\theta$. In particular, we will exploit the following exponential barrier function having $\lambda=10$ and $\alpha=100$, i.e., 
\begin{equation}
    p(\widehat{x}_k) \doteq 10e^{100(\hat{x}_k - 1)} + 10e^{100(-\hat{x}_k)}.
    \label{eqn:bf_exp}
\end{equation}

Let us consider a case with true parameter $\bar \theta = 3.5$, and $\widehat \theta = 3.95 +\mathcal{N}(0,10^{-4})$ as the current estimation of the identification algorithm, aiming to identify the true value. This is a ``critical" area for the parameters, very close to the values of instability ($\theta>4$), and in the region where the system exhibits chaotic behavior. Therefore, it is likely for the gradient, when no barrier functions {were} used, to update the estimated parameter in the direction of a possible local minimum near the values of $\theta>4$, causing gradient explosion. 
\begin{figure}[!tb]
    \centering
    \includegraphics[width = \columnwidth,]{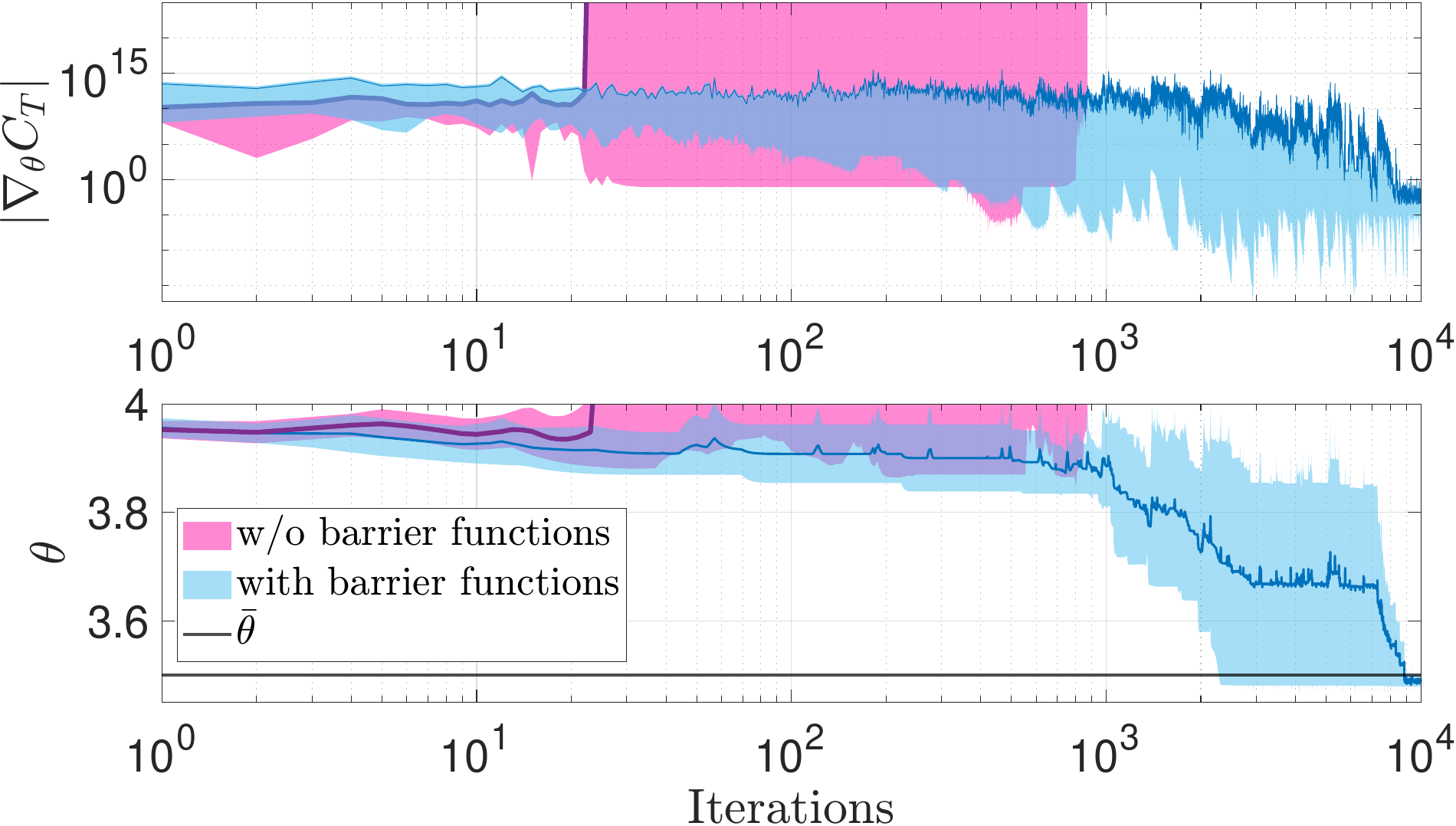}
    \caption{Effect of barrier functions on mitigating exploding gradients for $N=50$ different system initial conditions, represented with $\pm1$ standard deviation bands around the mean trajectories.}
    \label{fig:bf_example}
\end{figure}
Fig. \ref{fig:bf_example}  shows the evolution of the estimated parameters along with the computed gradient for $N=50$ randomly selected system initial conditions $x_0 \sim \mathcal{U}(0,1)$, when no barrier functions are used, and when the barrier function proposed in \eqref{eqn:bf_exp} is adopted in the loss function \eqref{eqn:final_cost}. One can notice that, when no barrier functions are used, the gradient explodes when the estimated parameter reaches the finite-time instability zone, i.e., $\widehat{\theta} > 4$. On the other hand, the use of a barrier function \eqref{eqn:bf_exp} helps the gradient to avoid updating the parameters to values that cause instability, until reaching a neighborhood of the true parameter value.

\section{Conclusions}\label{sec:concl}
In this paper, we introduced a novel and scalable method for solving nonlinear multi-step identification problems by computing exact gradients and leveraging first-order optimization techniques. By modeling the gradient evolution via a linear, time-varying dynamical system, we achieve exact gradient calculation by efficiently exploiting automatic differentiation.
The proposed approach 
allows a detailed analysis of gradient behavior throughout the optimization process, establishing conditions to avoid the exploding gradient phenomenon. Future works will focus on further refining the method and exploring its applicability to a wider range of systems, such as distributed and interconnected systems, comparing it to a broader class of identification methods in the literature. 

\bibliography{ifacconf}             

\appendix 
\section{Proofs}
\subsection{Proof to Proposition \ref{prop1}}
\label{app:prop1}
Consider \eqref{eqn:final_cost}. The gradient with respect to $\theta$ can be computed as
$
    \nabla_{{ \theta}}\mathcal{C}_{T} \doteq \frac{\d\mathcal{C}_T}{\d{\theta}} = \frac{\d}{\d{\theta}} \left( \sum_{k=1}^{T} \mathcal{L}_k \right) = \sum_{k=0}^{T} \frac{\d\mathcal{L}_k}{\d{\theta}}.
$
It follows that
\begin{equation}
    \nabla_{{\theta}}\mathcal{C}_{k} = \nabla_{{\theta}}\mathcal{C}_{k-1} + \frac{\d\mathcal{L}_k}{\d{\theta}},
    \label{eqn:cost_upd}
\end{equation}
where 
$
\nabla_{{\theta}}\mathcal{C}_{k} \doteq \sum_{\tau=0}^{k} \frac{\d\mathcal{L}_\tau}{\d{\theta}}
$ and $ \nabla_{{\theta}}\mathcal{C}_{0} = \mathbf{0}_{n_\theta} $.
Exploiting the chain rule of differentiation we have the following relation
\begin{equation}
    \begin{aligned}
        \frac{\d\mathcal{L}_k}{\d{ \theta}}^\top &=  
        {\frac{\partial \mathcal{L}_k}{\partial {\theta}}}^\top+
        \frac{\partial \mathcal{L}_k}{\partial e_k}^\top 
        \frac{\partial e_k}{\partial z_k}
        \frac{\partial z_k}{\partial {x_k}}
        \frac{\d {x_k}}{\d {\theta}}\\
    \end{aligned}
    \label{eqn:gradient_par}
\end{equation}
where the last term is defined as
\begin{equation}
    \frac{\d {x_k}}{\d {\theta}} = \frac{\partial {x_k}}{\partial {\theta}} + \frac{\partial {x_k}}{\partial {x_{k-1}}}\frac{\d x_{k-1}}{\d {\theta}}.
    \label{eqn:influence}
\end{equation}
%
Introducing the recursive memory operator $\Lambda_{k}$ defined in \eqref{eqn:Lambdadef}, we can rewrite \eqref{eqn:influence} as
$
    \Lambda_{k} = 
    \frac{\partial {x_k}}{\partial {\theta}} + \frac{\partial {x_k}}{\partial {x_{k-1}}}
    \Lambda_{k-1} = 
    \mathcal{J}^{x\!/\!\theta}_k + \mathcal{J}^{x\!/\!x}_k
    \Lambda_{k-1},
$
which yields \eqref{eqn:GUlaw_mat}. Thus, \eqref{eqn:gradient_par} can be rewritten as
\begin{equation}
     \frac{\d\mathcal{L}_k}{\d{\theta}}^\top ={\nabla_{\theta}^\top \mathcal{L}_k} + 
        {\nabla_{e}^\top \mathcal{L}_k} 
        {\mathcal{J}^{e\!/\!z}_k}
        {\mathcal{J}^{z\!/\!x}_k}
        \Lambda^\top_k = \varrho^\top_k + \rho^\top_k\Lambda_k,
        \label{eqn:grad_upd}
\end{equation}
using \eqref{eqn:rho-var-def}.
Thus, \eqref{eqn:GUlaw_grad} is obtained by transposing and substituting \eqref{eqn:grad_upd} in \eqref{eqn:cost_upd}, concluding the proof. 
\subsection{Proof to Proposition \ref{prop2}} \label{app:prop2}
Following the same reasoning of Appendix \ref{app:prop1}, consider
\begin{equation}
    \nabla_{{x_0}}\mathcal{C}_{k} = \nabla_{{x_0}}\mathcal{C}_{k-1} + \frac{\d\mathcal{L}_k}{\d x_0}.
    \label{eqn:cost_upd_x0}
\end{equation}
Exploiting the chain rule we have
\begin{equation}
    \frac{\d\mathcal{L}_k}{\d{x}_{0}}^\top = 
    \frac{\partial \mathcal{L}_k}{\partial e_k}^\top 
    \frac{\partial e_k}{\partial z_k}
    \frac{\partial z_k}{\partial {x_k}}
    \frac{\d x_k}{\d {x}_{0}},
    \label{eqn:gradient_par_x0}
\end{equation}
where the last term is defined as
$
    \frac{\d x_k}{\d { x}_{0}} = 
    \frac{\partial x_k}{\partial {x_{k-1}}}
    \frac{\d x_{k-1}}{\d {x}_{0}}.
$
Thus, recalling \eqref{eqn:phi0def}, 
we can write
$$
    \Lambda_{0,k} = 
    \frac{\partial x_k}{\partial {x_{k-1}}}
    \Lambda_{0,k-1} = \mathcal{J}^{x\!/\!x}_k\Lambda_{0,k-1},
$$
which yields \eqref{eqn:GUlaw_initcond_mat}. Then, \eqref{eqn:gradient_par_x0} can be rewritten as
\begin{equation}
    \begin{aligned}
        \frac{\d\mathcal{L}_k}{\d x_0}^\top = 
        {\nabla_{e}^\top \mathcal{L}_k} 
        {\mathcal{J}^{e\!/\!z}_k} 
        {\mathcal{J}^{z\!/\!x}_k} 
        \Lambda_{0,k} = \rho^\top_k{\Lambda_{0,k}},
    \end{aligned}
    \label{eqn:grad_upd_x0}
\end{equation}
using \eqref{eqn:rho-var-def}.
Thus, \eqref{eqn:GUlaw_initcond} is obtained by transposing and substituting \eqref{eqn:grad_upd_x0} in \eqref{eqn:cost_upd_x0}.  
\subsection{Proof to Theorem \ref{rmk:compl_improv}}\label{app:compl_improv}
The proof follows from basic notions on matrix computation complexities \citep{golub2013matrix}. 
One iteration of \eqref{eqn:GULAW} involves, in order, the multiplication $m_1 = \mathcal{J}^{x\!/\!x}_k\Lambda_{k-1}$, the sum $\Lambda_k = m_1 + \mathcal{J}^{x\!/\!\theta}_k$, the multiplication $m_2 = \Lambda^\top_k\rho_k$, and the sum $\nabla_{{\theta}} \mathcal C_{k} = \nabla_{{\theta}} \mathcal C_{k-1} + m_2 + \varrho_k$ with complexities $\mathcal{O}(n_x^2n_\theta)$, $\mathcal{O}(n_xn_\theta)$
$\mathcal{O}(n_xn_\theta)$, $\mathcal{O}(2n_\theta)$, respectively. Thus, the overall complexity for one iteration of \eqref{eqn:GULAW} is $\mathcal{O}(n_x^2n_\theta + 2n_xn_\theta + 2n_\theta) = \mathcal{O}(n_x^2n_\theta)$. It follows that, for $T$ iterations, the required complexity is~$\mathcal{O}(Tn_x^2n_\theta)$. 

\subsection{Proof to Theorem \ref{thm:exp.grad}}\label{app:exp.grad}
We observe that \eqref{eqn:GUlaw_grad} can be seen as a linear, time-varying (LTV) state-space system
with states $\nabla_{{\theta}} \mathcal C_{k} \in \mathbb R^{n_\theta}$, and
\begin{equation}
\begin{aligned}
A_k &= A \doteq I_{n_\theta} \in \mathbb R^{n_\theta,n_\theta},\, B_k \doteq (\Lambda^\top_k\rho_k + \varrho_k) \in \mathbb R^{n_\theta},\\
C_k &= C \doteq I_{n_\theta} \in \mathbb R^{n_\theta,n_\theta},\, D_k = D \doteq \mathbf{0}_{n_\theta} \in \mathbb R^{n_\theta},
\end{aligned}
\label{matrixLTV}
\end{equation}
with constant (bounded) input $u_k = 1 \in \mathbb R$, $\forall k \in [0,T]$.
By analyzing the BIBO stability properties of the LTV system defined by \eqref{matrixLTV}, we can study the exploding gradient phenomenon and obtain conditions under which the multi-step gradient remains bounded, i.e., condition \eqref{eqn:nexpdef} is satisfied.
The input-output behavior of \eqref{eqn:GUlaw_grad} is
specified by the unit-pulse response
$$
G(k,j)=C_k\Phi(k,j+1)B_j,\,k\geq j+1$$
with $\Phi(k,j)$ the transition matrix \cite[Chapter 20]{rugh1996linear},
defined for $k>j$ as
$$
\Phi(k,j) = \left\{
\begin{array}{ll}
     A_{k-1}A_{k-2}\dots A_j & k\geq j+1\\
     I & k=j .
\end{array}
\right.
$$
Stability results are characterized in terms of boundedness properties of $G(k, j)$.
From \cite[Theorem 27.2]{rugh1996linear} we have that the linear state equation~\eqref{eqn:GUlaw_grad} is uniformly BIBO stable if and only if there exists a finite constant $\mu_2$ such that the unit-pulse response satisfies
$
\sum_{i=j}^{k-1}\|G(k,i)\|\leq \mu_2
$
for all~$k$, $j$ with $k\geq j+1$.
Notice that for our system defined by \eqref{matrixLTV}, we have 
$
\Phi(k,j)=I_{n_\theta}$, $\forall k,j$, $G(k,j)=B_j$, $k\geq j+1,$
that yields~\eqref{eqn:theo2cond}, considering the multi-step horizon $[0,T]$.
\subsection{Proof to Corollary \ref{cor:corollary.exp.grad}} \label{app:corollary.exp.grad}
Considering \eqref{eqn:theo2cond} on the finite time interval $k \in [0, T]$, it is worth noting that the condition holds if $\Lambda_i$, $\rho_i$, $\varrho_i$ are bounded for all $i \in [j, k-1]$, for all possible $k \in [1, T]$ with $k\geq j+1$. 
The problem is hence reduced to understanding under which conditions $\Lambda_i$, $\rho_i$, $\varrho_i$ may not be bounded. Here, it follows from Lipschitz continuity that the functions $\rho_k$, $\varrho_k$ \eqref{eqn:rho-var-def} are always bounded since $\mathcal{L}_k(\cdot)$, $e_k$, $h(\cdot)$ are continuously differentiable. 
On the other hand, the evolution over time of $\Lambda_k$ is given by \eqref{eqn:GUlaw_mat}. Here, \eqref{eqn:GUlaw_mat} can be seen as a linear, time-varying state-space system 
with states $x_k = \text{vec}{(\Lambda_{k})} \in \mathbb R^{n_xn_\theta}$, input $u_k=\text{vec}{(\mathcal{J}^{x\!/\!\theta}_k)}\in \mathbb R^{n_xn_\theta}$, $A_k=I_{n_\theta} \otimes \mathcal{J}_k^{x\!/\!x}\in \mathbb R^{n_xn_\theta,n_xn_\theta}$, where $\text{vec}(\cdot)$ denotes the vectorization of a matrix, i.e., the column vector obtained by stacking the columns of the matrix on top of one another. Here, notice that the stability of this system is directly related to the matrix $\mathcal{J}_k^{x\!/\!x}$, corresponding to the linearization of the nonlinear system under analysis around the predicted trajectory obtained with the current values of~$\widehat \theta$ and $\widehat x_0$. This links the condition \eqref{eqn:theo2cond} directly to the stability of the predicted trajectories of the system (see, e.g., \cite[Chapter 22]{rugh1996linear}). It follows directly that, for the gradient to be bounded, the predicted trajectory must not diverge to infinity within the selected finite multi-step horizon interval.
\end{document}